\documentclass[aps,prd,twocolumn,groupedaddress,preprintnumbers,nofootinbib]{revtex4}
\usepackage{graphicx,epstopdf,amssymb,amsmath,verbatim,hyperref}
\pdfoutput=1

    \newcommand*{\F}{\mathbb{F}}
\newcommand*{\cy}{CY }
    \newcommand*{\httv}{H^{2,2}_{\rm vert}} 
    \newcommand*{\gsm}{G_{\rm SM}}

\DeclareMathOperator{\SU}{\mathrm{SU}}
\DeclareMathOperator{\SO}{\mathrm{SO}}
\DeclareMathOperator{\U}{\mathrm{U}}

\newcommand{\kobe}[1]{{{\textbf{[KL:\ #1]}}}}
\newcommand{\wati}[1]{{{\textbf{[WT:\ #1]}}}}

\newcommand{\clean}{
\renewcommand{\kobe}[1]{}
\renewcommand{\wati}[1]{}
}
\clean

\begin{document}

\preprint{MIT-CTP/5372}

\title{Natural F-theory constructions of Standard Model structure from $E_7$ flux breaking}

\author{Shing Yan Li}
\email{sykobeli@mit.edu}

\author{Washington Taylor}
\email{wati@mit.edu}

\affiliation{Center for Theoretical Physics,  Massachusetts Institute of Technology, Cambridge, MA 02139, USA}

\date{\today}

\begin{abstract}
We describe a broad class of 4D F-theory models in which an $E_7$
gauge group is broken through fluxes to the Standard Model gauge
group.  
These models are ubiquitous in the 4D F-theory
landscape and can arise from flux breaking of most models with
 $E_7$ factors.  
While in many cases the $E_7$ breaking leads to exotic matter, there
are large families of models in which the Standard Model
gauge group and chiral matter representations are obtained
through an intermediate $\SU(5)$ group.
The number of generations of matter appearing in these models
can easily be
small.
We demonstrate the possibility of getting three generations of
chiral matter as the preferred matter content.
\end{abstract}

\maketitle

\section{Introduction}

To describe the real world with string theory as a unified theory, it
has been a long-standing and primary goal to find
the structure of the Standard Model (SM) of
particle physics in string theory. In particular, we would like to
identify the SM as a \emph{natural} solution to string
theory. F-theory \cite{VafaF-theory,MorrisonVafaI,MorrisonVafaII}, a
strongly coupled version of type IIB string theory, is a particularly
promising framework for this purpose as it gives a global description
of a large connected class of theories (See \cite{WeigandTASI} for a
review).  F-theory gives 4D low-energy supergravity models when
compactified on elliptically fibered Calabi-Yau (CY)  fourfolds, which conveniently encode non-perturbative
brane physics into geometrical language. Combined with flux data, the
gauge symmetries and chiral matter content of any model can be easily
determined. Moreover, F-theory is dual to many other types of string
compactifications (such as heterotic).  We focus here on a novel class
of F-theory models that naturally give the SM gauge group and
chiral matter content.

There have been many attempts to build models with the SM gauge group
$G_{\rm SM}=\SU(3)\times \SU(2)\times \U(1)/\mathbb{Z}_6$ in
F-theory.  Starting from
\cite{Donagi:2008ca,BeasleyHeckmanVafaI,BeasleyHeckmanVafaII,DonagiWijnholtGUTs},
F-theory grand unified theories (GUTs) have been constructed, using
gauge groups of $\SU(5)$
\cite{Blumenhagen:2009yv,Marsano:2009wr,Grimm:2009yu,KRAUSE20121,Braun:2013nqa},
$\SO(10)$ \cite{Chen:2010ts}, etc. (See \cite{HeckmanReview} for
review) Recently, $10^{15}$ explicit solutions of directly tuned
$\gsm$ 
 were found in the string landscape \cite{CveticEtAlQuadrillion}. It
has also been argued that the SM matter representations generically
appear when $\gsm$ is directly tuned
\cite{TaylorTurnerGeneric,Raghuram:2019efb}. These results signal that
a considerable portion of the landscape may contain SM-like models.

These models cannot be the most generic or natural SMs in
the landscape, however. All the preceding gauge groups arise from
fine-tuning many moduli. In contrast, most F-theory
compactification bases have strong
curvature that enforces rigid (a.k.a\ {\it geometrically
  non-Higgsable} \cite{MorrisonTaylor4DClusters}) gauge symmetries,
which are present throughout the whole branch of moduli space
\cite{TaylorWangMC,HalversonLongSungAlg,TaylorWangLandscape}.
Furthermore, on many bases these rigid gauge factors forbid tuning additional factors like  $\gsm$.

A generic SM in the landscape can arise more naturally from the
geometric rigid gauge symmetries than through tuning moduli.  The rigid
gauge groups containing $\gsm$ are $E_8, E_7, E_6$, but not most other
traditional GUT groups \cite{MorrisonTaylor4DClusters}.
(While the non-abelian $\SU(3) \times\SU(2)$ of $\gsm$ can
arise as a rigid structure \cite{GrassiHalversonShanesonTaylor},
including the abelian factor is much more subtle
\cite{MartiniTaylorSemitoric, WangU1s}).  
In 4D, it  seems that  of these rigid GUT groups,
$E_8$ appears the most frequently in the landscape, while
$E_7$ and $E_6$ are also quite abundant 
\cite{TaylorWangMC,HalversonLongSungAlg,TaylorWangLandscape}. 
While $E_6$ has been one of the traditional GUT groups, little
attention has been paid to $E_7$ since it does not support chiral
matter.  We find here that, nevertheless, SM-like solutions can be
realized in F-theory by breaking rigid (or even non-rigid, tuned)
$E_7$ models. 

An economic way to tackle the above issues in F-theory is to turn on
$G_4$ flux inside a larger rigid group.
This can break the larger group
down to  $\gsm$, while inducing chiral matter in the broken
gauge group. In this paper, we describe F-theory models with
rigid $E_7$ and $G_4$ flux that leads to SM gauge group
and chiral matter spectrum with minimal supersymmetry.
Compared with other SM-like constructions in the past, our models
have the following novelties:

    $\bullet$ These models can be built using generic bases.
    Little fine-tuning is required to get the desired
    gauge group and chiral matter spectrum. They are thus
    more natural in the landscape.
    
    $\bullet$ Gauge groups with no chiral matter like $E_7$ can
    also be used as GUT groups. Non-complex representations
    in the unbroken gauge group can contribute to chiral
    matter in the broken group.
    
    $\bullet$ Chiral exotics are easily avoided even when we
    start with large GUT groups, although the 
 models we have
    identified without exotics involve an intermediate SU(5).

    $\bullet$ The resulting chiral multiplicities can easily be
    very small. It is natural, and sometimes preferred, to
    have 3 generations of chiral matter.
    
Although we focus here on $E_7$, there is a similar
construction for the similarly abundant rigid $E_6$. The
generalization is nontrivial since $E_6$ itself already
supports chiral matter.

The rest of this paper is organized as follows. We first
discuss some general features of 4D F-theory
compactifications. We describe the geometry of the rigid
$E_7$ model without assuming a specific base. 
We then review  vertical and remainder fluxes, and 
the flux
constraints that lead to consistent solutions. 
We show how models
with SM gauge
group and chiral matter spectrum can arise from a combination of
vertical breaking to SU(5) and hypercharge breaking from remainder
flux.
We give a simple explicit example of vertical flux
breaking to SU(5)  with
3
generations of chiral matter as the preferred
matter content. We conclude with some remarks and
future directions.

The arguments presented in this paper are minimal and aim at
describing our new class of SM-like models succinctly. We leave the more
general formalism and various technical subtleties to a longer
followup \cite{E7Long}

\section{$E_7$ gauge groups in F-theory}

A 4D F-theory model is defined by
 an elliptically fibered
\cy fourfold $Y$ over a threefold base $B$; this can be considered as
a non-perturbative type IIB string compactification on $B$. 
An $E_7$ gauge factor arises in the 4D supergravity theory when $Y$
is described by a certain form of Weierstrass model 
\cite{Kodaira,Neron,BershadskyEtAlSingularities}. Treating the elliptic curve as the
\cy hypersurface in $\mathbb{P}^{2,3,1}$ with
homogeneous coordinates $[x:y:z]$, $Y$ is given by the
locus of
\begin{equation}
y^2 = x^3 + s^3 f_3xz^4 + s^5 g_5 z^6 \,,
\label{eq:e7}
\end{equation}
where $s, f_3, g_5$ are functions on $B$ (more technically,
sections of line bundles ${\cal O} (\Sigma),{\cal O} (-4K_B-3
\Sigma),{\cal O} (-6K_B-5 \Sigma)$, with $K_B$ the canonical class of $B$)
and the seven-brane locus
$\Sigma$
 supporting the $E_7$ factor  is
given by $s=0$. There is adjoint matter $\mathbf{133}$ on the bulk of
$\Sigma$. There is also fundamental matter $\mathbf{56}$ localized on
the curve $s= f_3=0$, or $C_{\mathbf{56}}=-\Sigma \cdot (4K_B+3\Sigma)$
in terms of the intersection product, when the curve is nontrivial in
homology.

$E_7$ gauge factors can either be tuned by hand in the Weierstrass
model (\ref{eq:e7}) or can be forced from the geometry of $B$.
When a divisor (algebraic codimension-1 locus) $\Sigma$ on $B$ has a
sufficiently negative normal bundle $N_\Sigma$, singularities of the
elliptic fibration are forced to appear on $\Sigma$ so that any
elliptic fibration over $B$ automatically takes the restricted form
(\ref{eq:e7}), and there is a rigid (geometrically non-Higgsable)
$E_7$ gauge factor supported on $\Sigma$
%
\cite{MorrisonTaylor4DClusters}. 

The conditions for a rigid $E_7$ factor
 are satisfied for a large set of typical F-theory bases.
For 6D F-theory models, the toric bases have been completely
classified \cite{MorrisonTaylorToric} and 60\% of the
61539 allowed F-theory bases
have rigid $E_7$ factors. 
The total number of toric bases for 4D F-theory models is
${\cal O} (10^{3000})$ \cite{HalversonLongSungAlg,TaylorWangLandscape}, which is too large for explicit analysis.
A Monte Carlo estimate on a subset of these bases
(those without $E_8$
factors or codimension-2 $(4,6)$ singularities) gives roughly 18\%
with rigid $E_7$'s \cite{TaylorWangMC}, although the fraction
for all bases may be smaller
(the analogous subset for 6D bases contains 24483 bases of which
75\% have rigid $E_7$ factors). Similar
statistics may also apply to non-toric bases, but this question has not been addressed
in the literature.

\section{Fluxes and Gauge Symmetry Breaking}

\begin{figure}[t]
\begin{centering}
\includegraphics[width=0.8\columnwidth]{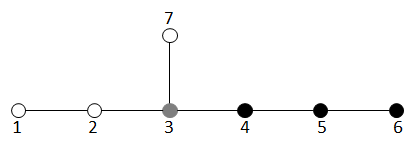}
\end{centering}
\caption{The Dynkin diagram of $E_7$. The Dynkin node labelled
$i$ corresponds to the exceptional divisor $D_i$. The solid
nodes are the ones we break to get $\gsm$. Node 3 (in gray) is broken by
remainder flux while the others are broken by vertical flux.}
\label{dynkin}
\end{figure}

The elliptic fibration
$Y$ with an $E_7$ factor over $\Sigma$ is singular. We need to consider
its resolution $\hat Y$  to study flux breaking using
 cohomology and
intersection theory on $\hat{Y}$. The resolution results in exceptional divisors
$D_i,1\leq i\leq 7$, corresponding to the Dynkin nodes of
$E_7$ (Fig.\ \ref{dynkin}).
The divisors $D_I$  on $\hat Y$ are spanned by the zero section
($z=0$) $D_0$, pullbacks of the base divisors $\pi^* D_\alpha$ (which
we also call $D_\alpha$ depending on context), and exceptional
divisors $D_i$ \cite{shioda1972,Wazir}. Note that while the choice of
resolution is not unique, our analysis and results are manifestly
resolution-independent \cite{Jefferson:2021bid}.

To break the $E_7$ factor we first turn on {\it vertical} $G_4$-flux
(see e.g. \cite{WeigandTASI}).
This lives in the space of $(2,2)$-forms spanned by products of
harmonic $(1,1)$-forms (which are Poincar\'e dual to divisors
$[D_I]$):
\begin{equation}
    \httv(\hat Y,\mathbb C)=\mathrm{span}\left(H^{1,1}(\hat Y,\mathbb C)\wedge H^{1,1}(\hat Y,\mathbb C)\right)\,.
\end{equation}
We expand  $G_4^\mathrm{vert}=\phi_{IJ} [D_I]\wedge[D_J]$ and work
with flux parameters $\phi_{IJ}$. We 
denote integrated flux as
\begin{equation}
    \Theta_{IJ}=\int_{\hat Y} G_4\wedge [D_I]\wedge[D_J]=\int_{\hat Y} G_4^\mathrm{vert}\wedge [D_I]\wedge[D_J]\,.
\end{equation}
We then have the resolution-independent relation
\cite{Grimm:2011fx,Jefferson:2021bid}
\begin{equation} \label{thetatophi}
    \Theta_{i\alpha}=-\Sigma\cdot D_\alpha \cdot D_\beta C^{ij} \phi_{j\beta}\,,
\end{equation}
where $C^{ij}$ is the Cartan matrix of $E_7$.

While $E_7$ can be broken
directly to $\gsm$ by vertical flux, this generally produces exotics.
To obtain models with only chiral SM matter, we also turn on the
following form of {\it remainder} flux \cite{Buican:2006sn,BeasleyHeckmanVafaII,Braun:2014xka}:
\begin{equation}
    G_4^\mathrm{rem}\in\mathrm{span}\left(\left[D_i|_{C_\mathrm{rem}}\right]\right)\,,
\end{equation}
where $C_\mathrm{rem}$ is a curve on $\Sigma$ but becomes homologically
trivial on $B$.
Some
non-toric bases have rigid $\Sigma$
with such curves, so that $\Sigma$ supports both rigid
$E_7$ and the {\it remainder} flux
\cite{Braun:2014pva,E7Long}.

$G_4$ satisfies certain constraints. To
preserve Poincar\'e symmetry we need
$\Theta_{0\alpha}=\Theta_{\alpha\beta}=0$ \cite{Dasgupta:1999ss}; this
condition is unaffected by $\phi_{i \alpha} \neq 0$. 
The flux
quantization condition is \cite{Witten:1996md}:
\begin{equation}
    G_4+\frac{1}{2}c_2(\hat Y)\in H^{2,2}(\hat Y,\mathbb R)\cap H^4(\hat Y,\mathbb Z)\,,
\end{equation}
where $c_2(\hat Y)$ is the second Chern class of $\hat Y$. We
will choose $Y$ with even $c_2(\hat Y)$ and consider the simple case where
$\phi_{IJ}$ is integral.  To preserve
supersymmetry we require primitivity \cite{Becker:1996gj,Gukov:1999ya}:
\begin{equation}
    J\wedge G_4=0\,,
\end{equation}
where $J$ is the K\"ahler form of $\hat Y$. This condition
stabilizes some K\"ahler moduli when the gauge group is
broken by vertical flux. Not all choices
of gauge-breaking
flux can stabilize $J$ within the K\"ahler cone, however. 
Finally we
have the D3-tadpole condition \cite{Sethi:1996es}:
\begin{equation}
    \frac{\chi(\hat Y)}{24}-\frac{1}{2}\int_{\hat Y}G_4\wedge G_4=N_{D3}\in\mathbb{Z}_{\geq 0}\,,
\end{equation}
where $\chi(\hat Y)$ is the Euler characteristic of $\hat Y$.
In general, $h^{2,2}>2\chi(\hat Y)/3\gg\chi(\hat Y)/24$. If
if we randomly turn on
flux in the whole middle cohomology such that the tadpole constraint
is satisfied, a generic flux
configuration vanishes or has small magnitude in most of the
$h^{2,2}$ independent directions.

We can now identify fluxes that break the model into $\gsm$  with SM chiral matter. 
If
$\Theta_{i\alpha}\neq 0$ for some roots
 $i$, the corresponding gauge bosons
become massive; similarly the appropriate linear combinations of Cartan
gauge bosons get masses through the St\"uckelberg
mechanism \cite{Grimm:2010ks,Grimm:2011tb}. 
%
%
To get the SM gauge group and exact chiral matter spectrum, we proceed
in two steps.  First we break (uniquely up to $E_7$
automorphism)
$E_7$ to $\SU(5)$ by turning
on vertical flux
with $\Theta_{i'\alpha}\neq 0$ for $i'=4,5,6$
(see Fig. \ref{dynkin}). This can be done by
turning on appropriate $\phi_{i\alpha}$ using Eq.
(\ref{thetatophi}). 
In the second step, in parallel with earlier work on tuned SU(5) GUT
models \cite{Donagi:2008ca,BeasleyHeckmanVafaI,BeasleyHeckmanVafaII,DonagiWijnholtGUTs,Blumenhagen:2009yv,Marsano:2009wr,Grimm:2009yu,KRAUSE20121,Braun:2013nqa},
we also turn on a remainder hypercharge flux
\begin{equation}
    G_4^\mathrm{rem}=\left[D_Y|_{C_\mathrm{rem}}\right]\,,
\end{equation}
where $D_Y=2D_1+4D_2+6D_3+3D_7$ is the exceptional divisor corresponding to the
hypercharge generator. This 
 breaks $\SU(5)$ to $\gsm$. Both $\mathbf{56}$ and
$\mathbf{133}$ are then broken into
SM matter:
\begin{equation}
    (\mathbf{3},\mathbf{2})_{1/6}\,,\quad
    (\mathbf{3},\mathbf{1})_{2/3}\,,\quad
    (\mathbf{3},\mathbf{1})_{-1/3}\,,\quad
    (\mathbf{1},\mathbf{2})_{1/2}\,,\quad
    (\mathbf{1},\mathbf{1})_{1}\,,
\end{equation}
along with an exotic $(\mathbf{3},\mathbf{2})_{-5/6}$ from
$\mathbf{133}$, which is non-chiral since it directly descends from the $\SU(5)$ adjoint.

The above {\it vertical} flux also induces chiral matter. To
calculate the multiplicities, we first
need to locate the matter surfaces (the
fibration over matter curves). A weight $\beta$ in a representation $R$
of $E_7$ can be expressed in the basis of simple roots $\alpha_i$:
\begin{equation}
    \beta=-\sum_i b_i\alpha_i\,.
\end{equation}
When localized on a matter curve $C_R$, we can decompose
the matter surface $S(\beta)$ into \cite{Bies:2017fam}
\begin{equation}
    S(\beta)=S_0 (R)+\sum_i b_i \left.D_i\right|_{C_R}\,,
\end{equation}
where $S_0 (R)$ only depends on $R$ and does not contain any
$\left.D_i\right|_{C_R}$ components. Since $E_7$ itself does
not support any chiral matter, $S_0 (R)$ does not contribute
to chiral multiplicities in the broken gauge group.

We now turn to the broken gauge group. Each set of values of
$b_{i'}$ for fixed $R$ give an irreducible representation $R'$ in the broken gauge group,
while the weights in $R'$ are spanned by $\alpha_i$ for
unbroken $i$. Different $R$ and different sets of
$b_{i'}$ can give the same $R'$, however. The chiral multiplicity of
$R'$ is then (generalizing \cite{Braun_2012,Marsano_2011,KRAUSE20121,Grimm:2011fx})
\begin{equation} \label{chiral}
    \chi_{R'}=\sum_R \sum_{b_{i'}(R)}\int_{S(\beta)} G_4\,,
\end{equation}
which can be easily computed using group theory data and
$\Theta_{i\alpha}$, thanks to the absence of $S_0 (R)$.
For $R=\mathbf{133}$ which lives on the
bulk of the gauge divisor $\Sigma$ instead of
a matter curve, the chiral multiplicity can be
computed by replacing $C_R$ with the canonical class
$K_\Sigma$ \cite{Bies:2017fam}, where $K_\Sigma=\Sigma \cdot (K_B+\Sigma)$
by adjunction. The above formula ensures that
$\chi_{R'}$ computed from different weights in $R'$ are all
the same. Moreover, we have $\chi_{R'}=-\chi_{\bar{R'}}$, and
anomaly cancellation is guaranteed \cite{E7Long}. In particular, the
above flux constraints imply
$\chi_{(\mathbf{3},\mathbf{2})_{-5/6}}=0$ regardless of
solutions.

\section{Small number of generations}


This class of SM-like models,
combining vertical and remainder fluxes, 
can be 
realized on a large class of
bases with rigid (or tuned) $E_7$ but cannot be
constructed completely in
simple toric geometries. Here for simplicity, to illustrate the
multiplicity of generations, we focus on {\it vertical} flux 
breaking to SU(5)
and give
an oversimplified example, using the Hirzebruch surface $\mathbb{F}_1$ as
the gauge divisor $\Sigma$.\footnote{This
oversimplification also leads to exotic $\U(1)$
gauge factors along with the $\SU(5)$
\cite{E7Long}. Here as a mere demonstration, we
focus on the $\SU(5)$ representations and ignore
the $\U(1)$ charges when calculating chiral
indices.}
When
further breaking to $\gsm$ through remainder flux is possible 
on more complicated surfaces,
this further breaking does not affect multiplicities.

$\mathbb{F}_1$ is a $\mathbb{P}^1$-bundle over another
$\mathbb{P}^1$. We denote $S$ as the section $\mathbb{P}^1$
and $F$ as the fiber $\mathbb{P}^1$. Then the intersection
numbers are $F^{2}=0,F\cdot S=1,S^{2}=-1$. Its anticanonical
class is $-K_\Sigma=2S+3F$. Now embed $\mathbb{F}_1$ into $B$ with normal bundle
$N_\Sigma=-aS-bF$. Let $F_S,F_F$ be divisors with
$\Sigma\cdot F_S,\Sigma\cdot F_F$ being pushforwards of
$S$ and $F$ into $B$ respectively. Without {\it remainder}
flux, we can assume $\Sigma\cdot F_S,\Sigma\cdot F_F$ are independent.
By choosing $N_{\Sigma}=-8S-7F$,
\begin{align}
    F_k=-4K_\Sigma+(4-k)N_\Sigma\,,\nonumber\\
    G_l=-6K_\Sigma+(6-l)N_\Sigma\,,
\end{align}
are both effective only when $k\geq 3,l\geq 5$, so we have a rigid $E_7$
supported on $\Sigma$ \cite{MorrisonTaylor4DClusters}. The nonzero
intersection numbers are then
$\Sigma\cdot F_S\cdot F_F=1,\Sigma^2\cdot F_F=-8,\Sigma\cdot
F_S^{2}=-1,\Sigma^2\cdot F_S=1,\Sigma^3=48$.

We claim that all the above constraints on {\it vertical}
flux can be solved inside the
K\"ahler cone by turning on nonzero but sufficiently small integer
$\phi_{iF_S}$ and $\phi_{iF_F}$ with opposite signs.
We require the
ratio $\phi_{iF_S}/\phi_{iF_F}$ to be the same for all $i$. To break
the gauge group, we turn on integer
$\phi_{5F_S},\phi_{6F_S}$ freely and
\begin{equation}
    \left(\phi_{1F_S},\phi_{2F_S},\phi_{3F_S},\phi_{4F_S},\phi_{7F_S}\right)=(2,4,6,5,3)n_S\,,    
\end{equation}
and similarly for $\phi_{iF_F}$, where
$n_S,n_F$ are integers with opposite signs. Eq. (\ref{chiral})
then gives a simple formula for the number of
generations of $\SU(5)$ GUT matter:
\begin{equation}
    \chi_{\mathbf{10}}=-\chi_{\mathbf{5}}=-7n_S-4n_F\,.
\end{equation}
This is a linear Diophantine equation and the number of
generations can be any sufficiently small integer.
As explained above, it is natural
to consider small $\phi_{iF_S}$ and $\phi_{iF_F}$. The
minimal flux configuration has $n_S=-1$ and
$n_F=1$, hence $\chi=3$ appears to be preferred.
This is an example of an
F-theory model with exactly three
generations of chiral matter, with minimal
fine-tuning.

The above is the most general {\it vertical} flux we can turn on
given the flux constraints and conditions on
$\Theta_{i\alpha}$. All other $\phi_{i\alpha}$ are
equivalent to a combination of $\phi_{iF_S}$ and
$\phi_{iF_F}$ by homology relations.

The above construction can be easily generalized to
incorporate hypercharge flux by using more complicated $\Sigma$'s on
non-toric bases \cite{Braun:2014pva,Braun:2014xka}
(with different multiplicities in each case).
We provide such explicit constructions in \cite{E7Long}.

\section{Conclusion and Remarks}

Within the framework of F-theory compactifications, we have
described a large class of SM-like models with the right
gauge group and chiral matter spectrum. These can originate
from rigid $E_7$ gauge symmetry, which is ubiquitous in the
landscape. String theory methods allow us to go beyond the
limit of field theories and use $E_7$ as a GUT group.
Remarkably, a subset of these models prefer 3
generations of SM chiral matter.
 Although we lack 
 an exact
quantification, we
believe that these models are more generic than tuned
SM-like models in the landscape.

Some 
 remarks and future directions: 

    $\bullet$ Although we only give an oversimplified example of the models,
    the same construction giving SM gauge group and chiral
    matter can be done on most bases containing rigid (or tunable) $E_7$
    factors.
In general, for other local geometries
supporting $E_7$, the number of generations may be different. In
    some cases, the chiral multiplicity is a multiple of 
    integers other than 3 and $\chi=3$ is forbidden. In
    most cases, $\chi=3$ is still allowed and generically
    natural because of small $\phi_{i\alpha}$ and the
    Diophantine structure, but this may not be the most preferred
    chiral matter content. In special cases, $\chi$ is
    a multiple of 3 and $\chi=3$ is both the minimal and
    preferred matter content.
    
$\bullet$ A generic base has many other rigid gauge factors
    apart from $E_7$'s. We can apply our SM-like construction on
    one of the $E_7$'s, while other gauge factors can serve
    as hidden sectors such as dark matter
    \cite{MorrisonTaylor4DClusters,Halverson:2016nfq}.

$\bullet$ One interesting feature of this construction of $E_7$ breaking
is that it
relies intrinsically on non-perturbative physics of F-theory and does
not have any immediately obvious description in the low-energy field
theory. It would be interesting to understand the structure of these
models better from the low-energy and/or dual heterotic pictures.

    $\bullet$ We have chosen a subset of embeddings of $\gsm$ into
$E_7$ which lead to SM chiral matter. 
The root embedding of $\SU(3) \times \SU(2)$ is unique up to
automorphisms;
however,
    there are other 
 embeddings of the U(1) factor that give various
    kinds of exotic chiral matter. 
    
    
    $\bullet$ It is clear that a similar construction as above
    also works for $E_6$. To get SM gauge group and chiral
    matter, we can use the same breaking pattern as in Fig.
    \ref{dynkin} but without the rightmost node. Calculating chiral multiplicities becomes more nontrivial, however,
    since $E_6$ itself supports chiral matter. There are more
    flux parameters $\phi_{ij}$ to turn on, and
    the matter surface $S_0 (R)$ also contributes
    nontrivially. This generalization is done in \cite{E7Long}.
    
    $\bullet$ We have been working with $E_7$, while
    the most generic rigid gauge group supporting $\gsm$ is likely $E_8$. A
    rigid $E_8$ generically contains codimension-2 $(4,6)$
    singularities, however, which signal the presence of
    strongly coupled superconformal sectors
    \cite{HeckmanMorrisonVafa,Apruzzi:2018oge}, and cannot
    be analyzed using our formalism. If we apply
    the same formalism to rigid $E_8$
    without this kind of singularity, we
    can break it into $\gsm$, but surprisingly no
    chiral matter is induced. In particular, the F-theory
    geometry with the most flux vacua \cite{TaylorWangVacua}
    contains $E_8$ instead
    of $E_7$ and does not support our formalism. Meanwhile, there have
    been similar attempts working with rigid $E_8$ using
    other formalisms like E-string theory \cite{TianWangEString}.

    $\bullet$ We have been focusing on the chiral matter
    spectrum, while knowledge of the vector-like spectrum is
    required for analyzing the Higgs sector and
    avoiding the exotic $(\mathbf{3},\mathbf{2})_{-5/6}$. This requires explicit cohomology data
    from topologically nontrivial 3-form potential
    backgrounds \cite{Bies:2014sra,Bies:2021nje,Bies:2021xfh}.
    Such data are much harder to analyze than $G_4$-flux and
    are beyond the scope of this paper.
    
    $\bullet$ Comparing with other tuned SM-like or GUT models, the
    origin of Yukawa couplings in our models is less clear
    due to several reasons. First,
    matter on both the bulk of $\Sigma$ and curve
    $C_\mathbf{56}$ are involved. It was argued that the
    Yukawa couplings between
    three bulk fields on $\Sigma$ always vanish if $-K_\Sigma$ is effective \cite{BeasleyHeckmanVafaI}. In contrast, we
    see no obstruction to having SM Yukawa couplings between
    the Higgs on $\Sigma$ and chiral matter on $C_\mathbf{56}$
    \cite{BeasleyHeckmanVafaI}, but a rigorous construction is
    still lacking. Besides, the Yukawa couplings
    between matter localized on curves are usually extracted from
    codimension-3 singularities on the base. Instead, the
    codimension-3 singularity on $C_\mathbf{56}$ has
    degree $(4,6)$, which goes beyond the Kodaira
    classification and may not be simply interpreted
    as Yukawa couplings. These $(4,6)$ points are associated
    with non-flat fibers, which possibly encode extra flux
    backgrounds and strongly coupled (chiral) degrees of
    freedom \cite{Candelas:2000nc,Lawrie:2012gg,Jefferson:2021bid}. These will be studied in a future
    publication \cite{46}.

We hope to address some of these issues in future studies,
and some of them will be explored in the followup paper \cite{E7Long}.
With this large class of SM-like constructions, we hope to shed
some light on where our Universe sits in the string theory
landscape, and whether it is a \emph{natural} solution in the landscape.

\begin{acknowledgments}
We would like to thank Lara Anderson, James Gray, Patrick Jefferson,
Manki Kim and Andrew Turner for helpful conversations.
This work is supported by DOE grant DE-SC00012567.
\end{acknowledgments}

\bibliography{references}

\end{document}